# A universal negative group delay filter for the prediction of band-limited signals

Henning U. Voss

*Abstract*—A filter for universal real-time prediction of band-limited signals is presented. The filter consists of multiple time-delayed feedback terms in order to accomplish anticipatory coupling, which again leads to a negative group delay for frequencies in the baseband. The universality of the filter arises from its property that it does not rely on a specific model of the signal. Specifically, as long as the signal to be predicted is band-limited with a known cutoff frequency, the filter order, the only parameter of the filter, follows and the filter predicts the signal in real time up to a prediction horizon that depends on the cutoff frequency, too. It is worked out in detail how signal prediction arises from the negative group delay of the filter. Its properties, including stability, are investigated theoretically, by numerical simulations, and by application to a physiological signal. Possible control and signal processing applications of this filter are discussed.

## I. Introduction

Signal prediction is one of the most often encountered problems in signal processing, machine learning, and artificial intelligence. In the majority of cases, signal prediction is based on a prior analysis of the signal to be predicted, and a subsequent model fit. The fitting procedure can be off-line, such as in some neural network approaches, which use a learning and a test data set, or in real time, such as in adaptive filtering. It would be highly advantageous for many applications if the modeling step could be omitted, which would result in a universal predictor.

Here, a *universal predictor* of a signal is understood to be a predictor that does not depend on an underlying model of the signal [1]. A solution to the universal prediction problem is proposed for a wide class of signals generated by deterministic or stochastic systems. It is based on a filter with a negative group delay, which in a sense to be specified below shifts signal components backwards in time, thus enabling prediction. This mechanism is independent of the specific signal shape or signal origin, as long as the signal is confined to the baseband, i.e., from zero frequency to a certain cutoff frequency, is stationary, and has limited variance. The limited bandwidth imposes temporal correlations on the signal, which in turn enable prediction by negative group delay. These correlations do not need to be "learned" by the predictor, and the filter coefficients are fixed, thus one can speak of a universal negative group delay (UNGD) predictor. The magnitude of the group delay defines the prediction horizon, i.e., the time the output $y(t)$, here called a predictor, predicts the input $x(t)$ ahead of time.

In general, negative group delay of an input/output system causes the output signal to anticipate or predict characteristics of the input signal. Negative group delay and the closely related concept of negative group velocity have been found in systems with anomalous dispersion [2-5], metamaterials [6-8], transmission lines [9, 10], and electronic circuits [11-15]. Recently, it has been shown that negative group delay can also occur in continuous-time systems with time-delayed feedback, or mathematically, non-autonomous delay-differential equations [16]. Time delays are a typical component of biological neuronal networks, and it is reasonable to hypothesize a possible relevance of delay-induced negative group delay in neuronal computations [17] involved, for example, in human motor control [18].

In order to understand group delay and negative group delay in particular, the input/output systems are typically analyzed by using frequency response or transfer function methods, an approach that will be pursued here, too. Because some of the concepts involved in this analysis, such as stationary phase approximation, the relationship between frequency response and cross-correlation function, and negative group delay, are often not well known, the presentation is kept in an informal tutorial style.

The outline of this article is as follows: In part II, the UNGD filter is defined and investigated without reference to any actual signals. First, it is specified as a fixed set of coefficients that only depend on the filter order but not the signal to be predicted. Then, the filter's stability properties are analyzed and it is shown that it is stable for any filter order. The prediction properties of the UNGD filter are quantified both in the time domain by the lag of the cross-correlation function and in the frequency domain by the filter's group delay. It is described how negative group delay and prediction are related, and how to determine the only free parameter, the model order. In part III, the filter's properties are demonstrated by the application to specific signals. In the first example, a simulated signal is used to test the theoretical predictions made in part II. Two counterexamples that violate the general requirement of band-limitation provide insight into possible limitations. The second example consists of an ECG signal to demonstrate applicability to real-world data. Part IV consists of a discussion about potential application fields and how the UNGD filter relates to previous work.

## II. THE UNGD FILTER

### A. Definitions

Given a stationary, zero-mean, variance- and band-limited signal $x(t)$ ($t = \ldots, -\Delta t, 0, \Delta t, 2\Delta t, \ldots$), the UNGD filter is given by

$$y(t) = bx(t) - \sum_{k=0}^{m-1} c_k y(t - (m-k)\Delta t) \quad (1)$$

with real filter coefficients $b$ and $c_k$, defined as

$$b = \frac{3+m}{2}, \quad c_k = \frac{k+1}{m} \quad (k = 0, \ldots, m-1). \quad (2)$$

These coefficients and the filter order $m \geq 2$ completely specify the UNGD predictor. For example, for $m = 3$ one has

$$y(t) = 3x(t) - \frac{1}{3}y(t - 3\Delta t) - \frac{2}{3}y(t - 2\Delta t) - y(t - \Delta t). \quad (3)$$

For the sake of simplicity, the sampling time $\Delta t$ will be set to unity and omitted in the following. The filter output is supposed to predict the signal in the following sense:

$$y(t) \approx x(t + \delta), \quad (4)$$

where $\delta > 0$ is the *prediction horizon*. In other words, the output of the filter is a real-time estimate of future signal values.

In the remainder of Section II it will be numerically demonstrated and analytically proven that the UNGD predictor is indeed a stable predictor for low-frequency band-limited signals. The rationale behind the particular choice of filter coefficients (2) is as follows: The coefficients $c_k$ are weights for the delayed terms $y(t - (m-k))$ and decrease linearly with increasing delay time. (Note that $k = 0$ corresponds to the largest delay, $m$). In general, the further delayed a term in Eq. (1), the more it contributes to possible instability of the filter. Therefore, in order to keep the filter stable, terms with large delay need to be weighted less than terms with a small delay. The coefficient $b$ sets the filter gain to unity for zero frequency. There are probably many possible configurations for the choice of coefficients, and here only one possibility, Eq. (2), is considered. Whereas it is not guaranteed that this particular choice defines an optimal universal predictor in terms of prediction horizon and accuracy, it defines a stable predictor with a reasonable prediction horizon.

The frequency is defined up to the Nyquist frequency, i.e., $\omega \in [0, \pi]$ for the angular frequency and $f \in [0, 1/2]$ for the frequency $f = \omega/2\pi$. Low-frequency band-limited signals are defined as signals with significant power only in the baseband $[0, f_0]$. The filter order $m$ will be determined by the signal cutoff frequency below. Roughly, the lower the cutoff frequency, the higher the filter order and the larger the prediction horizon. Prediction performance will be quantified a-posteriori by the *cross-correlation function* (CCF) between filter output and time-shifted signal values, i.e.,

$$C_{xy}(\tau) = \frac{E[(x_\tau - \mu_{x_\tau})(y - \mu_y)]}{\sigma_{x_\tau}\sigma_y}, \quad (5)$$

where $E$, $\mu$, and $\sigma$ denote the expectation value, mean and standard deviation, respectively, and $x_\tau(t)$ is defined as $x(t - \tau)$. The actual estimation of the CCF from two time series of discretely sampled data usually is being performed via fast Fourier transforms and provided in the supplementary code (https://codeocean.com/2017/10/16/universal-negative-group-delay-lpar-ungd-rpar-filter/). The prediction horizon $\delta$ is defined as that lag that maximizes the CCF in the sense that

$$\delta = \arg\max C_{xy}(-\tau) \quad (\tau > 0). \quad (6)$$

The prediction horizon is always positive. Should the CCF have a global maximum of its magnitude for positive delays, there is no prediction but lag and $\delta$ is not defined. This convention will become useful later on when the argument of the maximum of the CCF is identified with the group delay of the filter, which is negative when the filter is predictive.

For real-time prediction, no future time points with respect to the reference time $t$ can be used. Whereas the filter (1) has been written already in causal terms, there remains to re-write the prediction equation (4). Due to signal stationarity, one can perform a time shift on both sides of Eq. (4) to obtain the equivalent causal prediction expression

$$y(t - \delta) \approx x(t). \quad (7)$$

In this causal expression, the filter output is anticipating the filter input. How the filter (1) accomplishes this is the topic of this article.

### B. Filter stability

The UNGD filter (1), (2) will be analyzed by its frequency response function

$$H(\omega) = \frac{b}{1 + \sum_{k=0}^{m-1} c_k e^{-i(m-k)\omega}}. \quad (8)$$

It defines the input/output relationship between $x(t)$ and $y(t)$ under steady-state conditions in Fourier space as

$$Y(\omega) = H(\omega)X(\omega), \quad (9)$$

where $X(\omega)$, $Y(\omega)$ are the Fourier transforms of $x(t)$ and $y(t)$, respectively [19, 20]. The stability of the filter is determined by the location of the roots of the denominator with respect to the complex unit circle [21, 22]. Due to the infeasibility to compute all the roots for general sets of coefficients, here a less general but sufficient condition for stability is used: The filter (1) is stable when the real part of the denominator of the frequency response function (8) is positive [23-25]. This can be simplified to a condition for the *stability spectrum*

$$S(\omega) = 1 + \sum_{k=0}^{m-1} c_k \cos((m-k)\omega) > 0 \; (\omega \in [0, \pi]). \quad (10)$$

This stability criterion can be tested easily for any filter order and corresponding set of coefficients. For the specific configuration of the UNGD filter given by Eq. (2), some example spectra are provided in Fig. 1 for various values of the filter order $m$.

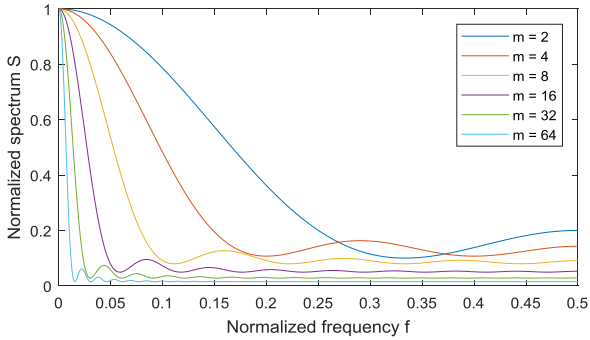

Fig. 1: The stability spectrum $S(f)$, Eq.(10), for various filter orders. The UNGD filter is guaranteed to be stable if the spectrum is positive for all frequencies, which is the case. The normalized frequency range is from zero to the Nyquist frequency ½. The spectra have been scaled to unity for zero frequency, by dividing it by $b$.

Stability of the UNGD filter with order $m \geq 2$ can be proven analytically: The stability spectrum is

$$S(\omega) = 1 + \sum_{k=0}^{m-1} c_k \, \text{Re}(e^{-i(m-k)\omega}) \; (\omega \in [0, \pi]), \quad (11)$$

or, with inserted coefficients from Eq. (2) and $z = e^{i\omega}$,

$$S(z) = 1 + \frac{1}{m} \text{Re} \, z^{-m} \sum_{k=0}^{m-1} (1+k) z^k. \quad (12)$$

For $z = 1$ ($\omega = 0$), one has $S(1) = \frac{m+3}{2} = b > 0$. For $z = -1$ ($\omega = \pi$), one has $S(-1) = \frac{m-1}{2m} > 0$ for odd $m$ and $S(-1) = \frac{1}{2} > 0$ for even $m$. For all other $z$, one can eliminate the sum in Eq. (12) by using the relationships

$$\begin{aligned}\sum_{k=0}^{m-1} z^k &= \frac{1-z^m}{1-z}, \\ \sum_{k=0}^{m-1} k z^k &= \frac{z^m((m-1)(z-1)-1)+z}{(1-z)^2}\end{aligned} \quad (13)$$

to obtain

$$S(z) = 1 + \frac{1}{m} \text{Re} \, \frac{z^{-m} + mz - m - 1}{(1-z)^2} \; (z \in (-1,1)). \quad (14)$$

With $z - 1 = e^{i\frac{\omega}{2}} \left(e^{i\frac{\omega}{2}} - e^{-i\frac{\omega}{2}}\right) = 2i e^{i\frac{\omega}{2}} \sin\left(\frac{\omega}{2}\right)$ follows $(z-1)^2 = -4 e^{i\omega} \sin^2\left(\frac{\omega}{2}\right)$, and with $\sin^2\left(\frac{\omega}{2}\right) = \frac{1-\cos(\omega)}{2}$ one obtains finally

$$S(\omega) = 1 - \frac{\cos(\omega(m+1)) - (m+1)\cos(\omega) + m}{2m(1-\cos(\omega))} \quad (15)$$
$$(\omega \in (0, \pi)).$$

It remains to be shown that this analytically given stability spectrum is larger than 0 for all $\omega \in (0, \pi)$ (i.e., with $\omega = 0, \pi$ excluded): From the last expression and the requirement $S(\omega) > 0$ one can derive the inequality $\cos(\omega(m+1)) < m + (1-m)\cos(\omega)$. It is clear that the left hand side never exceeds 1, and the right hand side is always larger than 1, so the inequality is fulfilled. This concludes the proof of the stability of the UNGD filter.

Remark: By proving positivity of the stability spectrum it was also shown that the UNGD filter has a strictly positive real frequency response function [26], with the usual implication that it can be synthesized by discrete ideal passive linear elements.

One might wonder whether the filter could not be simplified in general by using fewer coefficients. That this is not the case is demonstrated in Fig. 2.

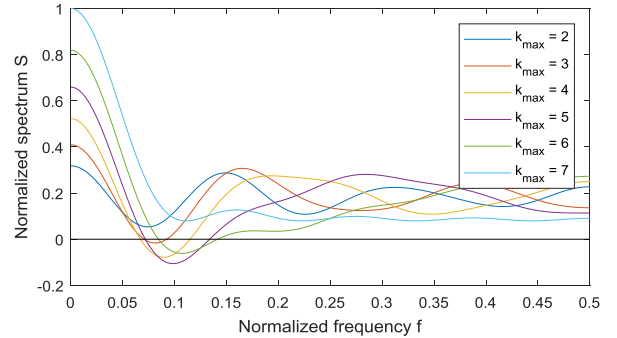

Fig. 2: The stability spectrum $S(f)$, scaled as in Fig. 1, for the filter with order $m = 8$, in which only the first coefficients are used, up to $k_{max}$, and the rest are set to zero. It is clear that the spectrum is not indicating guaranteed stability for some cases in which not all coefficients are being used, corresponding to graphs with values smaller than zero. Adding the last coefficient ($k_{max} = 7$) then guarantees stability.

## C. Prediction properties - cross correlation function

In the following, the theory of the UNGD predictor is outlined for the continuous-time case, in which both time and delay times are continuous variables. The application to discrete-time data would follow directly by sampling the data, for example by multiplication of the continuous-time signal with a Dirac comb [27]. Numerical computation of all the quantities defined below from sampled data is provided in the supplementary code, specified before.

The CCF can be written using the cross power spectrum $P_{XY}(\omega) = X(\omega)Y^*(\omega)$ as

$$C_{xy}(\tau) = \frac{1}{\sigma_x \sigma_y} \int_{-\pi}^{\pi} P_{XY}(\omega) e^{i\omega\tau} d\omega. \quad (16)$$

Defining the auto-power spectra $P_{XX}(\omega) = X(\omega)X^*(\omega)$ and $P_{YY}(\omega) = Y(\omega)Y^*(\omega)$, with the relation (9) it follows

$$H(\omega) = \frac{P_{XY}(\omega)}{P_{XX}(\omega)} = \frac{\sigma_y f_{XY}}{\sigma_x f_{XX}}, \quad (17)$$

where $f_{XY} = \frac{P_{XY}}{\sigma_x \sigma_y}$ and $f_{XX} = \frac{P_{XX}}{\sigma_x^2}$ are the normalized cross- and auto-power spectral densities, respectively. Therefore,

$$C_{xy}(\tau) = \frac{1}{\sigma_x \sigma_y} \int_{-\pi}^{\pi} P_{XX}(\omega) H(\omega) e^{i\omega\tau} d\omega. \quad (18)$$

The output signal variability can be computed as

$$\sigma_y^2 = \int_{-\pi}^{\pi} Y(\omega) Y^*(\omega) d\omega = \int_{-\pi}^{\pi} P_{XX}(\omega) |H(\omega)|^2 d\omega$$
$$= \sigma_x^2 \int_{-\pi}^{\pi} f_{XX}(\omega) |H(\omega)|^2 d\omega. \quad (19)$$

Therefore, the correctly normalized CCF results as

$$C_{xy}(\tau) = \frac{\int_{-\pi}^{\pi} f_{XX}(\omega) H(\omega) e^{i\omega\tau} d\omega}{\sqrt{\int_{-\pi}^{\pi} f_{XX}(\omega) |H(\omega)|^2 d\omega}}. \quad (20)$$

This completes the specification of the CCF for the UNGD predictor (1), (2). It can be easily computed numerically for given frequency response functions (8) and arbitrary, even non-integer, delays, as provided in the supplementary code.

A special case arises for band-limited signals $x$ with uniform power up to the cutoff frequency $\omega_0$. In this case the auto-power spectral density of the input signal is

$$f_{XX}(\omega) = \frac{1}{2\omega_0} \text{ for } \omega \in [-\omega_0, \omega_0], \quad (21)$$
$$= 0 \text{ otherwise,}$$

and the CCF simply becomes

$$C_{xy}(\tau) = \frac{\int_{-\omega_0}^{\omega_0} H(\omega) e^{i\omega\tau} d\omega}{\sqrt{\int_{-\omega_0}^{\omega_0} |H(\omega)|^2 d\omega}}. \quad (22)$$

In summary, the relationships derived in this section provide a complete description of the expected prediction properties as described by the CCF for low-frequency band-limited signals with uniform power spectrum.

## D. Group delay

An alternative way to describe the prediction properties of the filter is its group delay. The frequency-dependent group delay quantifies how much each signal frequency component is shifted backwards in time [28]. For example, for a constant negative group delay over a certain frequency range, all signal components in this range will be shifted the same amount, and if the signal is band-limited to this range, the signal will be predicted in real time [16]. In this section, this will be formalized for model (1), (2). It will turn out that the group delay for zero frequencies is determined alone by the filter order. This group delay extends to frequencies larger than zero, as has been shown for similar filters before [16-18].

The group delay of the UNGD filter is defined over the filter's phase. The frequency response function generally can be written in terms of phase and gain as

$$H(\omega) = G(\omega) e^{i\Phi(\omega)}. \quad (23)$$

The gain determines amplification/attenuation of the signal. In particular, the coefficient $b$ given in Eq. (2) follows from the condition that the gain $G(0) = 1$ in order to avoid attenuation or amplification of the signal at zero frequency. The filter's phase delay is

$$\tau_p(\omega) = -\frac{\Phi(\omega)}{\omega}, \quad (24)$$

and its group delay is

$$\tau_g(\omega) = -\frac{d\Phi(\omega)}{d\omega}. \quad (25)$$

Back to the UNGD. Setting

$$\beta(\omega) = \left|1 + \sum_{k=0}^{m-1} c_k e^{i(m-k)\omega}\right|^2, \quad (26)$$

one can define the only phase-relevant terms $R$ and $I$ via

$$R + iI = \frac{H(\omega)\beta(\omega)}{b} \quad (27)$$

$$= 1 + \sum_{k=0}^{m-1} c_k \cos((m-k)\omega) + i \sum_{k=0}^{m-1} c_k \sin((m-k)\omega),$$

such that the phase and group delays in terms of $R$ and $I$ are

$$\tau_p(\omega) = -\arctan\left(\frac{I}{R}\right)/\omega \tag{28}$$

and

$$\tau_g(\omega) = -\frac{d}{d\omega}\arctan\left(\frac{I}{R}\right) = -\frac{R^2}{R^2+I^2}\frac{d}{d\omega}\frac{I}{R}, \tag{29}$$

respectively.

Of particular relevance for low-frequency band-limited signals is the behavior of the group delay for the case of zero frequency. It is

$$\tau_g(0) = -\frac{\sum_{k=0}^{m-1} c_k(m-k)}{1+\sum_{k=0}^{m-1} c_k}. \tag{30}$$

For the set of coefficients (2) it becomes

$$\tau_g(0) = -\frac{m^2+3m+2}{3(m+3)} = -\frac{m}{3} - \frac{2}{3m+9} \\ \approx -\frac{m}{3} \text{ for large } m. \tag{31}$$

The phase delay for small frequencies has a similar behavior as the group delay but cannot explain prediction as defined by means of the CCF. The next paragraph provides the reason.

*E. Negative group delay and prediction*

In the following, how negative group delay entails prediction is derived. In particular, it is worked out how the group delay of a filter determines the prediction horizon as defined over the maximum of the CCF (5), a discussion that is normally not easy to find in the literature (but see [29-31]). In addition, the prevailing notion is that negative group delay can only play a role for predicting the envelope of a signal (with exceptions including Refs. [11, 32, 33]). However, in numerical simulations below it will be demonstrated that arbitrary low-frequency band-limited signals can be predicted. Rather than presenting a rigorous approach, the goal is to provide an understanding of how negative group delay entails prediction.

One needs to show that generally the prediction horizon follows from the group delay of the filter. Towards this end, the frequency response function in Eq. (20) is split into gain and phase, and assuming unit variance of the input signal, one gets

$$C_{xy}(\tau) = \int_{-\pi}^{\pi} \frac{1}{\sigma_y} f_{XX}(\omega) G(\omega) e^{i\Phi(\omega)} \cdot e^{i\omega\tau} d\omega \\ = \int_{-\pi}^{\pi} \sigma(\omega) e^{i(\Phi(\omega)+\omega\tau)} d\omega. \tag{32}$$

We are interested in those delays that maximize the cross correlation function. The stationary phase approximation (see [31] and references therein) asserts that the most significant contributions to maximizing the integral are those for which the magnitude of the exponent is changing only little with frequency, as faster changing contributions approximately cancel out by destructive interference. A necessary requirement for this to happen is that the function $\sigma(\omega)$ changes only relatively slowly. The stationary phase approximation is demonstrated in Fig. 3 for the case of uniformly band-limited data, Eq. (22). It is evident from Fig. 3a that for the chosen parameters, which correspond to the numerical example below, the integrand with $\tau = -5$, shown as highlighted black graph, yields the highest value of the integral (32). Therefore, the CCF in Fig. 3b is maximized for this corresponding argument.

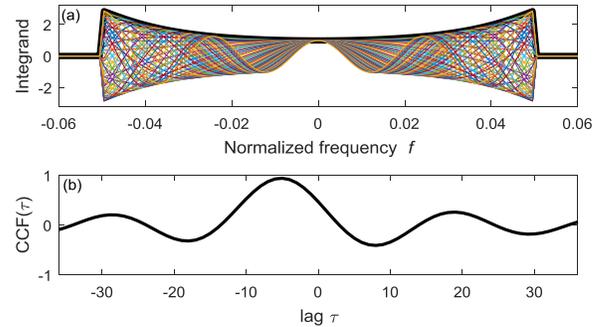

Fig. 3: Visualization of the stationary phase approximation. In (a), the real parts of the integrands of Eq. (32) are shown for lags $\tau = -36 \ldots 36$. The integrand for a lag of $\tau = -5$ is highlighted in black. Apparently, of all graphs this is the component that yields the highest value of all, and thus the CCF has its maximum at this particular lag, which is shown in (b). Parameters used in Eq. (32), via Eqs. (22) and (21), are $\omega_0/2\pi = f_0 = 0.05$ and $m = 18$. Band limitation of the noise signal was performed with a Butterworth filter of order 15.

To obtain phases for which the phase exponent in (32) is constant or stationary in frequency, the frequency derivative should vanish [31]:

$$\frac{d}{d\omega}(\Phi(\omega)+\omega\tau) = 0, \tag{33}$$

or

$$\frac{d\Phi(\omega)}{d\omega} = -\tau. \tag{34}$$

Therefore, the delay in this expression, which was coming from the definition (16) of the CCF via the cross power spectrum, is formally identical to the group delay (25). This way, the group delay is the proper definition for the prediction horizon measured by the CCF.

For the sake of completeness, it might be interesting to discuss again the *phase* delay and what it means for prediction. For the special case of $\Phi(\omega) + \omega\tau = 0$ in Eq. (32), the phase would be stationary, too, and the definition of the phase delay (24) would result. Indeed, for the case of low-frequency band-limited signals with a phase-offset of zero for zero frequency and otherwise linear phase behavior, the phase delay would yield the same signal shift. However, it does not indicate the correct frequency limits up to which the filter is predictive. This will be demonstrated in the numerical counterexample a) below. Also, in more general cases in which the phase has a finite offset for the frequency interval in question, the phase delay is not the correct description either. For example, for signals consisting of high- and separate low-frequency components (the previously mentioned "envelope" case), the phase offset normally is not zero. The group delay, but not the phase delay, definition then removes this offset and yields the correct signal shift.

*F. Specification of the filter order from the cutoff frequency of the signal*

As mentioned already, the only free parameter of the UNGD filter is the cutoff frequency $f_0$ of input signal. From the cutoff frequency follows the filter order $m$, from which again all filter coefficients follow via Eq. (2). The prediction horizon to be expected is provided by Eq. (6). In order to completely specify the filter for a given application, it thus remains to provide a rule for obtaining the filter order.

Numerical plots of the group delay show that it starts out for small frequencies at negative values and then at some frequency becomes positive (see Fig. 5). At this crossover point, for a filter with $m = 4$ the gain is $G = 2.8$. For much larger filter order, for example, $m = 40$, it is only slightly larger, $G = 3.1$. Therefore, in order to obtain a rule of thumb for the cutoff frequency, it is selected as that frequency for which the gain for the first time reaches a value of $G = 3$. Figure 4 shows the relationship between the cutoff frequency and the filter order. It can be used to determine the filter order from the observed data cutoff frequency, i.e., from that frequency that defines the interval that contains most of the signal power.

This concludes the specification of the filter parameters. The whole filter application procedure will be summarized in the examples to follow.

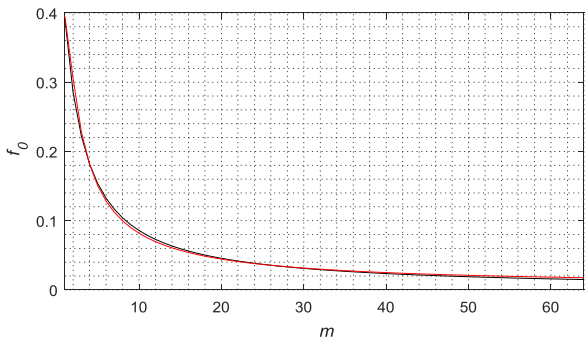

Fig. 4: Black graph: The cutoff frequency $f_0$ as a function of filter order $m$. This graph can be used to obtain an optimal filter order from the data cutoff frequency. Red graph: Alternatively, the relationship is approximated by the function fit $f_0 = -0.42\, m^{-2} + 0.81\, m^{-1} + 0.005$. The reason why the cutoff frequency is shown as a function of filter order and not the other way round, which could be more convenient, is that the inverse function fit seems to require more terms in order to be as accurate.

### III. EXAMPLES

*A. Numerical band-limited noise*

As an example for the application of the filter a numerical simulation is considered first. All simulations are programmed in MATLAB R2017a (The MathWorks, Inc., Natick, MA). The script for this example is made available as supplemental data.

The signal consists of 1024 points of smoothed white noise with zero mean. The smoothing is the same as used in Fig. 3, i.e., a Butterworth filter of order 15 with a cutoff frequency of 0.05. Figure 5a shows part of the signal in black. The signal's power spectrum estimate is shown in Fig. 5d, confirming the cutoff frequency at 0.05. Based on this cutoff frequency, a filter order of $m = 18$ is selected via the relationship provided in the caption of Fig. 4. This specifies all filter coefficients via Eqs. (2) and Fig. 4.

The filter output, or the predicted signal, is shown in Fig. 5a, too, in red. It is evident that the output signal has a negative phase shift towards the input signal, and is also slightly distorted. This distortion is expected in systems with negative group delay [14, 34] and can be understood by taking a closer look at the filter's behavior in frequency space:

The frequency response function (8) is shown in terms of its gain and phase in Fig. 5f, together with its estimate from data. This estimate results from Eq. (17), realized by using MATLAB's cross power spectral density function (cpsd, see supplementary code). For frequencies higher than the cutoff, due to lack of signal power, the estimates of phase and gain of the frequency response function become unreliable. Shown in red is also the theoretical group delay, defined by Eq. (29). It starts out at zero frequency with the value of $\tau_g(0) = -6.03 \approx -6$, Eq. (31), but then increases. Therefore, we would expect a prediction horizon $\delta$ of at most 6. Note that only the frequency response function of the model (1), (2) has been used to derive this result, no actual simulated data. Both the theoretically expected CCF, Eq. (22), and the estimated CCF, Eq. (5), shown in Fig. 5b, indicate a group delay of -5. The theoretical group delay for continuous time signals follows from Eq. (22) as -5.14. A linear fit to the approximately linear signal phase in the frequency range of [0, 0.05] in Fig. 5c yields an average slope of 4.8, yielding a group delay estimate of -4.8 via Eq.(25). These latter estimates of the group delay are smaller than the group delay for zero frequency due to the increase of the group delay over frequency, Fig. 5f.

The theoretical phase delay, defined by Eq. (24), is shown in green. It starts out matching the group delay but then stays negative for all frequencies up to the Nyquist frequency (outside of plot area). The filter gain starts out with a value of one, by definition via Eq. (2), and then increases, too. When it reaches a value of $\approx 3$, the group delay reaches its crossover point from where it becomes positive. The crossover point corresponds to the cutoff frequency defined by Fig. 4 and happens to be almost identical to the Butterworth filter cutoff frequency of 0.05 used to filter the white noise.

The prediction accuracy is quantified by the maximum value of the estimated CCF, which is 0.93. The maximum value of the theoretically predicted CCF, based on the filter's frequency response function, is 0.94. In this calculation via Eq. (22) it is assumed that the Butterworth filter yields a perfect signal frequency cutoff, which is not the case in reality, and that the signal is infinitely long. Numerical simulations with larger data sets show that the estimated and the theoretical CCF indeed can become visually indistinguishable. The prediction accuracy is also visualized in Fig. 5e as a phase portrait that plots signal and predicted signal values against each other. It is a graphical visualization of the relationship (6).

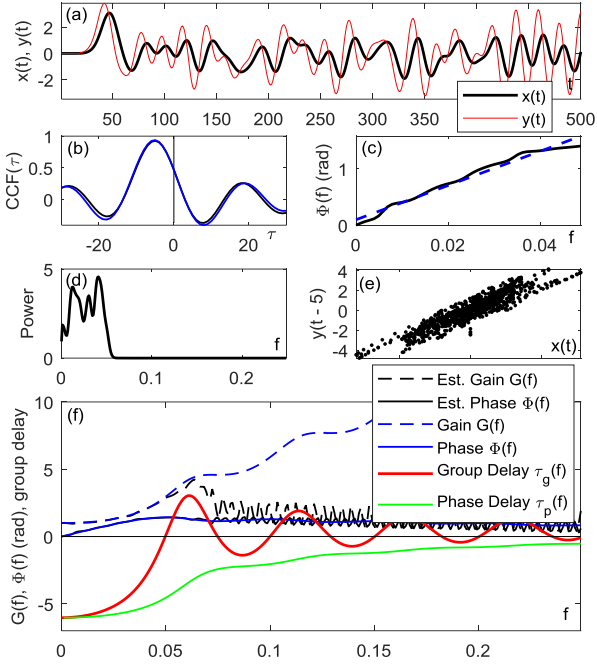

Fig. 5: Application of an order $m$ =18 filter to a signal consisting of band-limited white noise. (a) Input (black) and output time series (red). Only the first 500 of the 1024 data points are shown. (b) The theoretical (blue, Eq. (22)) and data CCF (black, Eq. (5)). Both attain their maximum at the same value of $\tau_{max}$ = -5. The vertical line marks $\tau$ = 0. (c) Fit of the group delay (blue) on the phase estimated from data (black), over the frequency range with significant signal power, i.e., up to the cutoff frequency of 0.05. The estimated group delay is $\tau_g$ = -4.8. (d) Signal power spectrum. (e) Phase portrait of $x(t)$ vs. $y(t - \tau_{max})$. (f) Theoretic frequency response function gain G and phase $\Phi$, Eq. (23), and their estimates from data, as well as the theoretical group delay $\tau_g$ from Eq. (25) and phase delay from Eq. (24). Estimates are accurate for frequencies below the cutoff only, due to the lack of out-of-band signal power. The maximum theoretic gain (outside of plot area) is 44.

The same example is used to study the effects of the filter order $m$ in Fig. 6. As mentioned before, the filter order is the only parameter to be determined before applying the filter, and all filter coefficients follow via Eqs. (2) and Fig. 4. Therefore, it is important to know what to expect for cases in which the filter order cannot be determined accurately beforehand. Examples include cases where only a limited amount of data exists before application of the filter, or signals that are not perfectly stationary, or signals that do not have a well-defined cutoff frequency. By systematically increasing the filter order from 1 to 40 and applying the filter to always the same signal, Fig. 6 shows how the prediction horizon and prediction accuracy are affected. From these graphs it is evident that choosing a larger than necessary filter order does not result in an improved prediction horizon, as it stays constant over a wide range, but in a degradation of prediction accuracy, indicated by a decreasing CCF. Conversely, selecting a too small filter order gives away the potential of a larger prediction horizon, but the prediction accuracy increases. Overall it can be concluded that the exact determination of the filter order is not too critical; a rule of thumb for applications in which the filter order can be adapted to the data beforehand could be to prioritize whether to optimize the prediction horizon or prediction accuracy, and then set the filter order higher than the number provided by Fig. 4 or lower, respectively.

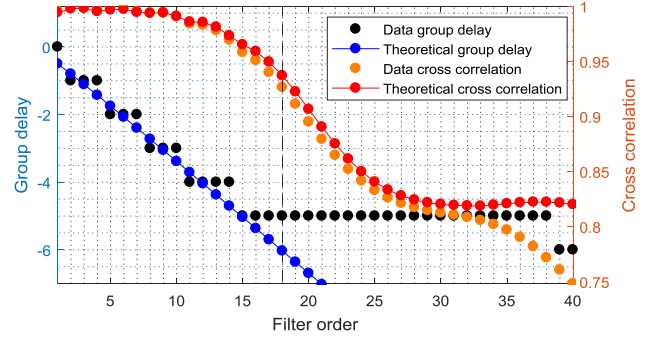

Fig. 6: Systematic study of group delay and prediction accuracy in dependence of filter order $m$ for the same data as used in Fig. 5. Blue bullets show the theoretical group delay at zero frequency, Eq. (31), for increasing filter orders $m$. These values can be non-integer numbers. Black bullets show the group delay estimated from input/output data, using the filter with order $m$ applied to always the same band-limited noise data set. These integer values equal the location of the maximum of the CCF (5). Orange bullets show the maximum CCF value estimated from data, Eq. (5), and red bullets the theoretical CCF (22). The filter order of $m$ = 18 used in Fig. 5 resulted from the equation in the caption of Fig. 4 with a cutoff frequency of 0.05 and is marked by the dashed vertical line.

### B. Counterexamples

In the following, the filter is used to demonstrate its behavior if the conditions on the data are not met.

a) The signal is not restricted to the baseband: For signals with significant power outside of the baseband, the filter might not be predictive. Figure 7 shows such an example in which the signal power is contained within the frequency band [0.05, 0.075]. This band corresponds to the first positive group delay band in Fig. 5. The other parameters are kept fixed. It can be seen that the signal is more amplified due to the higher gain (Fig. 7a) and is lagging the input (Fig. 7b): The magnitude of the CCF has its maximum at $\tau$ = 5 with a value of -0.97, indicating a positive group delay and anti-correlation. A linear fit to the estimated phase yields a group delay of 2.1. The theoretical *phase* delay is still negative (Fig. 5f) and thus does not describe the observed signal lag correctly.

b) The signal is inherently unpredictable: A band-limited signal to which sudden jumps are added is an example for being inherently unpredictable, as the jumps do not have deterministic precursors. After each jump, the filter is in a transient state, and a negative group delay is generally not to be expected during these times. The UNGD filter does not predict the jumps because the signal power spectrum includes out-of-band frequency components. This is demonstrated in Fig. 8.

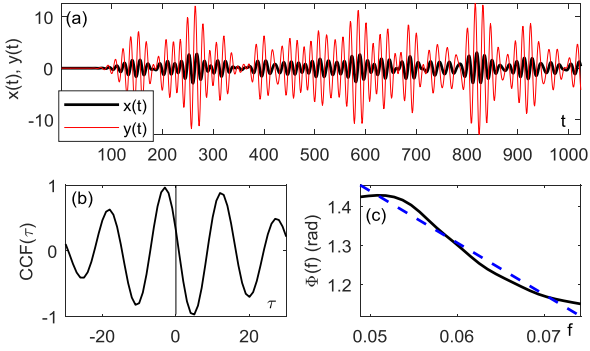

Fig. 7: Application of an order $m$ =18 filter to a signal consisting of band-limited white noise that is not restricted to the baseband and violates the assumptions necessary for prediction. (a) Input (black) and output time series (red). (b) The CCF estimated from data (black). It attains its magnitude maximum at $\tau_{max}$ = 5. The vertical line marks $\tau$ = 0. (c) Fit of the group delay (blue) on the phase estimated from data (black), over the frequency range with significant signal power. The estimated group delay is $\tau_g$ = 2.1.

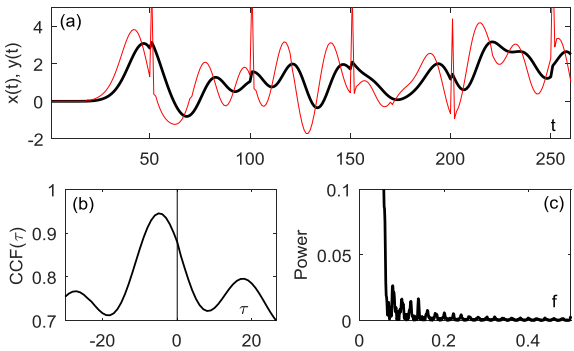

Fig. 8: The same data as in Fig. 5 but with an added upward jump of 0.5 each 50 data points. (a) Input (black) and output time series (red). It is evident that the sudden upward jumps are not predicted, although overall the filter is still predictive as shown by the CCF, which still attains its maximum at a negative delay of $\tau_{max}$ = -5. The added upward jumps violate the assumption that the signal is limited to the baseband, seen in the power spectrum in (c).

### C. An ECG signal

In the second example a publicly available electrocardiogram (ECG) trace from an intensive care setting is used to demonstrate feasibility of real world data applications and to test for the prediction of events, here the prediction of the 'R' peak of the ECG. ECG records and other parameters recorded in an intensive care setting are often low-pass filtered before digitization and thus band-limited. An ECG trace can be considered as a combination of stochastic and deterministic dynamics; whereas the QRS complex is a recurring pattern, its occurrence in time can be variable, depending on cardiac health status [35]. The first application of a negative group delay filter to predict band-limited physiological signals in real time has been performed by Dajani [32].

The ECG trace considered here was selected *ad hoc* as the first one from a public data base with 250 patients (mgh001.dat, physionet.org/physiobank/database/mghdb/, [36, 37]). Within this record, a segment of the first ECG electrode signal with stationary data, i.e., no apparent change of baseline signal, was selected, comprising eight QR waveforms. The mean signal value was subtracted. Otherwise the signal was not further processed. It contains a visible 60 Hz contamination component, corresponding to a normalized frequency of 0.17 (the signal sampling rate is 360 Hz). Based on the power spectrum, a cutoff frequency of 0.1 was selected, i.e., well below the 60 Hz contamination part, which falls into the high-gain part of the filter. This cutoff frequency corresponds to a filter order $m$ = 8 via Fig. 4. Again, the filter is completely specified by this parameter.

Figure 9 shows a comparison of the input and output signals of the filter at intervals centered around the R peaks. The R peaks are predicted, as well as sometime the small negative Q waveforms. Also, the 60 Hz contamination is amplified. The CCF is maximized at -2 samples with a value of 0.95. Theoretically, the zero frequency group delay is expected to be -2.7 samples, Eq. (31), and the theoretical CCF, Eq. (22), is maximized at -2 samples with a value of 0.94. However, in this example the signal power spectral density is not uniform and Eq. (22) can only be an approximation for the theoretical CCF.

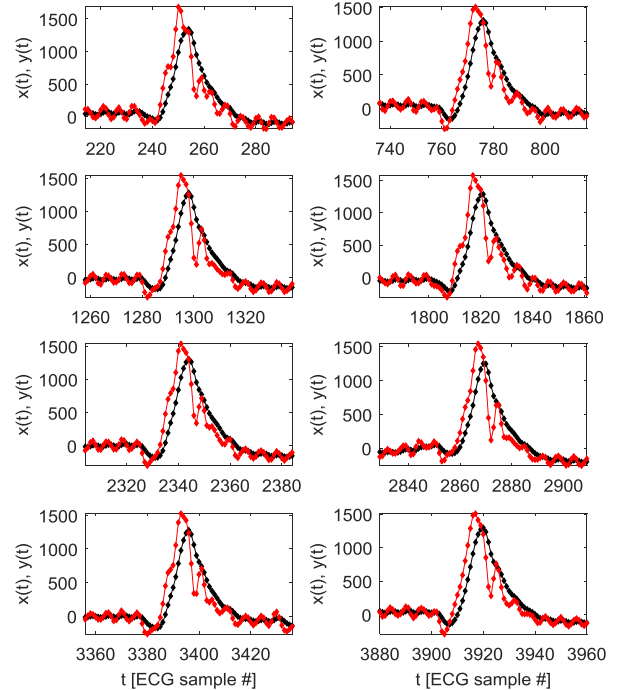

Fig. 9: Eight sections of an ECG signal centered around eight consecutive QRS complexes (black), and their predicted values (red).

The question arises if prediction can be improved if the ECG signal is cleaned from the 60 Hz contamination by off-line band-stop filtering. Although this is not real-time signal prediction anymore, it provides additional insights and is presented in Fig. 10. After removal of the signal contamination, the filter order can be increased to $m$ = 9 and the CCF has its maximum at -3 samples with a value of 0.97. Theoretically, the zero frequency group delay is expected to be -3.1 samples, and the theoretical CCF is maximized at -3 samples with a value of 0.92. Therefore, signal prediction could be improved, but also evident from Fig. 10 is that prediction of the QR waveform already starts with anticipation of the Q drop (in the first of the figures, this is the dip at about t = 240 samples). It is not evident by eye what part of the signals actually informs the filter that the Q dip is to be expected. Therefore, it would probably be very difficult to design a non-negative group delay-based filter to accomplish such a task. In summary, these results do not seem to be trivial, in particular under the consideration that the UNGD filter is universal.

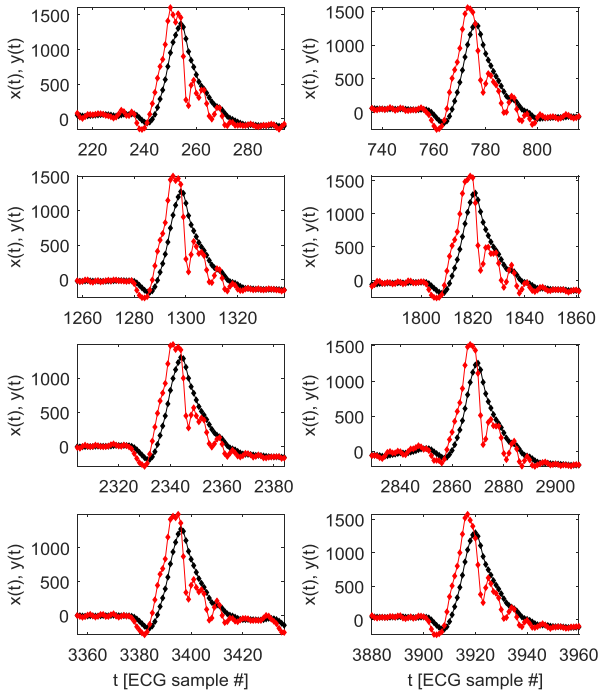

Fig. 10: Same as Fig. 10 but with a pre-filtered ECG signal for removal of the 60 Hz contamination. The prediction horizon is now 3 samples.

## IV. DISCUSSION

### A. Prediction without signal memory

The UNGD filter is a discrete-time filter, which allows for digital signal processing implementation. It is also a *delay-induced* negative group delay predictor, which has been described previously for continuous-time signals [16-18]. The delay-induced negative group delay concept differs from most other predictors or forecasting models, which do not rely on past predicted but past *input* signal values [38, 39]. In other words, most conventional time series predictors can be written in the form: predicted $x(t + \tau) = f(x(t), x(t-1), x(t-2), ...)$, in which $f(.)$ is a specific model of the time series to be predicted. These conventional predictors mostly rely on prior observations rather than prior predictions, and usually contain coefficients resulting from a fit to a fixed learning data set or which are continuously being updated [40]. In contrast, the UNGD predictor cannot be written in this explicit form, as it does not use past input signal values $x(t-1)$, $x(t-2)$, ... for prediction, but only previously predicted output values $y(t-1)$, $y(t-2)$, ..., along with the present input value $x(t)$. The previously predicted output values are delayed feedback inputs to the predictor. In this sense, the UNGD predictor is a pure *infinite impulse response* (IIR) filter. "Pure" refers to the property that it does not have a finite impulse response part, i.e., the counter of the frequency response function is just a constant. IIR filters have been shown to possess negative group delay before [41]. The pure IIR scheme could have advantages in natural [42, 43] or man-made [44] applications, as it does not require to store the input signal in memory but only the previously predicted values. Therefore, the translation process from a sensory input to the internal coding of information needs to happen only once, at the present time point, but not for signals restored from memory. Thinking of a nervous system, this translation happens for example in the retina while generating neuronal spikes, and no original sensory signals are stored in the brain. Further, the main ingredient for delay-induced negative group delay are time delays, which are abundant in the central nervous system. Real-time prediction by the pure IIR principle could be a significant advantage not only for living systems but also for artificial agents.

### B. Anticipatory coupling, robotics, and the neurosciences

A similar universal predictor, but for time-continuous signals, has been studied before; anticipatory relaxation dynamics (ARD) [16] is caused by delay-induced negative group delay, too [17], achieved via anticipatory coupling. Anticipatory coupling [16, 45-48] refers to the design that the input signal value $x(t)$ at the present time $t$ is compared or "coupled" to time delayed output values of the predicting system, $y(t - \tau)$. Whereas in ARD anticipatory coupling is more evident than here, because coupling terms like $x(t) - y(t - \tau)$ are used, in the UNGD filter multiple delayed feedback terms are used, each still defining anticipatory coupling. Therefore, it might be instructive to compare both approaches: Whereas the maximum prediction horizon of the UNGD filter is only one third of the maximum delay used in the filter, the ARD mechanism achieves a maximum prediction horizon of one half of the used delay. However, whereas both filters can be designed to be stable for any delay used, the UNGD filter allows for selecting larger feedback delays without losing the predictive capabilities and stability. One could also say that by adding more data into the "memory" of the filter [49], its performance increases.

With respect to the possible application to human motor control, the advantage of having multiple delays has recently been emphasized by Stepp and Turvey [50]. There, the closely related method of anticipatory synchronization [28, 46] is referred to, but the reliance on anticipatory coupling is identical to the ARD and UNGD filters. Multiple feedback delays also play an important role in artificial neural networks [51] but their potential for prediction via anticipatory coupling has not been used yet to the best of my knowledge (with the exception of Ref. [52]).

Whereas it is conceivable that ARD or related concepts arising from anticipatory synchronization could be "implemented" in the human brain at different levels of description [17, 18, 53-55], the question arises if the UNGD model with its multiple delayed feedbacks plays any role in the functioning of human motor control. Even if the maximum possible prediction horizons achieved here seem to be small, it has been shown that they can easily match experimentally observed anticipation times in human motor control [18, 56, 57]. From here on it becomes speculation, but if this is indeed the case, it would be quite likely that a neuronal realization of multiple feedback delays to accomplish anticipatory coupling need to include the cerebellum as a key component. Not only that the cerebellum plays a crucial role for fine motor control [58], but its peculiar neuronal architecture and function [59] might provide the necessary combination of complexity and regularity to accomplish a multiple delay feedback filter. In any case, if we can learn how nature solves predictive control problems [60-67], these insights might be valuable in robotic control [68, 69], too. This UNGD filter proposal could provide another contribution to the growing utilization of anticipatory coupling mechanisms towards robotics and other engineering applications [44, 45, 70-76].

### C. Adaptation

The only parameter of the filter, the filter order, determines the cutoff frequency of the band-limited input signals to be predicted. If this is not known a-priori, or if the signal is not strictly band-limited, adaptation to unknown signals could be accomplished as follows: It has been shown above that too large filter orders are not harmful in the sense of losing stability or sudden loss of predictability. Therefore, in order to adapt the filter to an unknown signal or signals that are not strictly band-limited, one could start with a large filter order and then reduce it until a useful compromise between possible prediction horizon and prediction accuracy has been reached.

### D. Filter stability and cascading

The UNGD filter is stable for all filter orders order $m \geq 2$. In choosing the filter coefficients as described here, well-known problems of IIR filter stability have been avoided. First, the stability of the UNGD filter does not depend on the filter input [22]. Second, it is not necessary to check the eigenvalues for each set of coefficients or to implement rescue mechanisms to guard against potential instability [77].

Since the filter is stable, feeding the filter output into the same filter again (cascading), again results in a stable combined filter. In general, cascading negative group delay systems might increase anticipation times at the cost of prediction accuracy and system stability [14, 15, 34]. One reason for reduced accuracy is that out-of-band frequency components are amplified in the first filtering step and then fed into the next filter application, becoming even more amplified. Numerical simulations (not shown) on the ECG data, for example, result in an increased prediction horizon of the cascaded filter, but also in a reduced correlation coefficient at the new prediction horizon. It is an open question if cascading in general will help to increase the prediction horizon or if it is more useful to use a larger filter order from the start.

### E. Causality

The UNGD filter is causal because in the time domain it only refers to present and past but not to future states. For spatially extended systems, things are more involved because of the time a signal travels, which involves group and phase velocities. In the realm of spatially extended systems, it has been shown that there are media that actually require a negative group delay for some frequencies in order to be, overall, causal [78], and again it is the group and not the phase delay that matters in these discussions. Furthermore, in media with strong gain for some frequencies, the group delay would be negative for zero frequency. And, in any medium, the group velocity is abnormal (i.e., greater than the vacuum speed of light, infinite, or negative) for a frequency at which the absorption has an absolute maximum [78]. Analogous behaviors can be observed in the examples presented here, in which the filter gain turned out to be smallest for frequencies with negative group delay. (Re-setting the constant $b$, which does not affect the group delay and has defined such as to set the gain to unity at zero frequency, might as well define signal attenuation at zero frequency.) Note that the theoretical group delay in Fig. 5 alternates between negative and positive values. This is a result of the causality restrictions on the filter's frequency response function. The phase delay stays negative for all frequencies in this example; it is not bound by causality constraints. Along these lines, Ref. [34] provides an analysis of how negative group delay in non-spatially extended systems is limited by causality. In summary, somewhat paradoxically one might say that the predictive properties of the UNGD and other negative group delay filters arise from their causality.